# Using radioactivities to improve the search for nearby radio-quiet neutron stars

Markus M. Hohle[1,2], Ralph Neuhäuser[1], Nina Tetzlaff[1]

1) Astrophysikalisches Institut und Universitäts-Sternwarte
Friedrich-Schiller-Universität Jena
&
2) Max-Planck-Institut für extraterrestrische Physik
Garching b. München

mhohle@astro.uni-jena.de

**Abstract**

Neutron stars (NS) and black holes (BH) are sources of gravitational waves (GW) and the investigation of young isolated radio-quiet NS can in principle lead to constraints of the equation of state (EoS). The GW signal of merging NSs critically depends on the EoS. However, unlike radio pulsars young isolated radio-quiet neutron stars are hard to detect and only seven of them are known so far. Furthermore, for GW projects it is necessary to confine regions in the sky where and of which quantity sources of GW can be expected. We suggest strategies for the search for young isolated radio-quiet NSs. One of the strategies is to look for radioactivities which are formed during a supernova (SN) event and are detectable due to their decay. Radioactivities with half lives of ~1 Myr can indicate such an event while other remnants like nebulae only remain observable for a few kyrs. Here we give a brief overview of our strategies and discuss advantages and disadvantages.

## 1. Introduction

Most of the 2000 known neutron stars are radio pulsars; however, there are seven objects which have been identified as young radio-quiet isolated neutron stars known as the so-called Magnificent Seven (M7, see Table 1). Multiplying the age of our Galaxy with the SN rate of about 1-3.3 events/100yr (Tammann et al., 1994) our Galaxy should host roughly $10^8$ NSs. The known radio pulsars may represent only a tiny fraction of the entire NS population in our Galaxy. However, the SN rate estimation by Tammann et al. is based on counting SNe in the 300 brightest galaxies. The apparent luminosity depends upon the inclination of the galaxies to the line of sight, the intergalactic absorption and the value of the Hubble constant. Recent estimates adopted from star forming rates by Reed (2005) lead to 1-2 SN events/Myr, but were extrapolated from within 1.5 kpc to the Galaxy.
For an overview of galactic SN rates obtained from different methods see Diehl et al. (2006).
Close-by (<600 pc) young (~Myr) NSs such as the M7 show pure thermal emission (see Haberl, 2007) and are detectable as faint objects in the optical band and as soft X-ray sources. They in principle enable us to constrain the EoS. The radius can be obtained from brightness, distance and temperature (optical, X-rays) using the Stefan-Boltzmann law (see e.g. Trümper et al., 2004) and the mass possibly from sub-stellar companions (but very rare, see e.g. Posselt et al. 2008), gravitational micro lensing (unlikely) or absorption lines.
However, distances have been determined for only two of the M7 (see Table 1): RX J1856.5-3754 and RX J0720.4-3125. Since the distance of the latter is only poor known and it is a variable source RX J1856.5-3754 seems to be the only object suitable for constraining the EoS owing to its accurate known distance although it does not show absorption features (see Braje & Romani, 2002 and Trümper et al., 2004). Proper motions have been obtained only for three members of the M7. During the next decades no micro lensing event is expected to occur for any of the three of them.
For those reasons enlarging the sample of M7-like objects could lead to enhance the understanding of such sources.

## 2. Search Strategy

While M7 like sources are hard to detect we suggest three strategies for a systematic search to find more of them:

1. To search for regions in the sky which host clusters of massive stars with a near-zero residual life time (i.e. age minus expected life time). Some members of these clusters may already have exploded and are now young and hot NSs or BHs.
2. To compare these regions to origins of runaway stars (see Prokhorov & Popov, 2002) and
3. to look for regions with a high amount of isotopes which are formed during a SN event, as a third evidence.

Those isotopes can be identified by their gamma ray emission due to their decay. Suitable isotopes are 26Al and 60Fe with decay times ~1Myr.
The age estimation for ordinary stars from luminosity and temperature depends upon assumed metallicity and the model used. To tracing back a runaway star to its non optical component we have to assume that first the pre SN orbit was circular, second the mass loss of the runaway star during the SN is negligible and third the non-optical component did not receive any kick-velocity. The last point decreases the probability of locating the non optical component to a few percent.
Using these three strategies in combination we have three evidences for certain regions in the sky where we can expect to find M7-like NS in a higher probability to observe them.
We used the Simbad catalogue to determine all stars within 3 kpc which are supposed to end in a SN, i.e. have masses of about 8 solar masses or more. These are stars earlier than B4 on the main sequence, earlier than B8 for luminosity classes IV and III and all spectral types for luminosity classes II and I, i.e. massive red giants and super giants.
Spectral types were obtained from Simbad, HIPPARCOS and CCDM (**c**atalogue of the **c**omponents of **d**ouble and **m**ultiple stars). We derived V magnitudes from HIPPARCOS and CCDM and B magnitudes from HIPPARCOS only (all corrected for multiplicity). To calculate extinctions from multi colour photometry we additionally used JHK from 2MASS (**t**wo **m**icron **a**ll **s**ky **s**urvey) in combination with intrinsic colours predicted by models by Bessell et al. (1998), Kenyon & Hartmann (1995) and Schmidt-Kaler (1982).
For most stars of the Simbad catalogue HIPPARCOS parallaxes and proper motions are available. Effective temperatures and bolometric corrections were obtained from spectral types and by means of tables from SK82/KH95.
Knowing distances, extinctions, bolometric corrections and magnitudes one can calculate luminosities. Together with luminosities and temperatures we can estimate masses and ages by putting the stars into the H-R diagram. Therefore we used evolutionary models from Bertelli et al. (1994), Schaller et al. (1992) and Claret (2004). All this is presented in Hohle & Neuhäuser (2008).
We calculated the median and standard deviation of the mass for each star for masses obtained from the models mentioned. The median of the standard deviation of the masses is less than 7.7% compared to the median masses themselves (assuming solar metallicity for all stars). We regard this as satisfactory consistency of the models.
With known age and mass it is in principle possible to estimate the residual life time for each star, i.e. the remaining time until the star ends in a SN. Life times were estimated using models from Kodama (1997), Tinsley (1980) and Maeder & Meynet (1989).
The procedure is shown in Figure 1.
In the case of the existence of several SN progenitors with near-zero residual life times (verified by applying different life time models) within a small volume (clusters) we can assume that all members either will explode within a few 0.1 Myrs or some unknown members have already exploded. We plan to improve the calculations using specific metallicities for OB associations.

Considering isotopes as evidence for NSs formed in star clusters see Voss et al., this volume, discussing the abundance of 26Al in the interstellar medium.

## 3. Results and Discussion

We estimate the future SN rate of the Gould Belt which is a star forming region surrounding us within 600 pc (see e.g. Neuhäuser, 1997 and Pöppel & Olano, 1981).

Applying a SN rate of 20-27 events/Myr for the Gould Belt (see Grenier, 2000) and assuming that 2/3 of all massive stars within 600 pc belong to the Gould Belt (Comeron & Torra, 1994) up to 50 M7-like NS which are younger than 1.5 Myr should exist in the solar vicinity.

From radial velocity and proper motions (if available) we identified 170-180 runaway stars (i.e. space velocities larger than 35 km/s) within 3 kpc with masses larger than 8 solar masses. Their number varies with the model applied for mass estimation. Tracing back every runaway star may lead us to the non-optical component which is supposed to be a NS or BH now. Runaway stars are shown in Figures 3 and 4.

Identifying a star as a runaway star depends on the reliability of its parallax. However, for HIPPARCOS parallaxes the error is as large as the value itself for distances from about 1 kpc. Nonetheless, HIPPARCOS parallaxes are applicable to such investigations as we restrict our search to close-by M7-like NSs (more distant ones are probability not detectable yet). Errors in parallax and proper motion affect the conclusion of which OB cluster the runaway originated. Furthermore, the uncertainties of distance influence the age estimation because of luminosity.

A typical size of a certain region where the non-optical component of a runaway star can be expected is relatively large (~5°x5°) considering all errors (see Prokhorov & Popov, 2005).

With the predicted age of a runaway star we have an upper limit for the age of the non-optical component. This age can be compared to the age of the association from which the runaway star originated as well as to cooling models for the NS.

Owing to the large uncertainties each of the two methods (runaway stars and clusters with near-zero life times) is not reliable individually.

If we compare the regions containing clusters of stars with near-zero residual life time to possible origins of runaway stars (OB associations of such regions are marked in Figure 2 and have typical sizes of 30'x30') and furthermore to regions with a relative large amount of isotopes like 26Al or 60Fe (formed in clusters of massive stars, which may host a young NS) we have three evidences to constrain areas in the sky which host more NSs and BHs than in average.

Absorption of gamma rays due to the interstellar medium is negligible within 600 pc. So we may identify an M7-like NS which we would miss if we only look for soft X-ray sources. Determining the origin of an expanding region of 26Al and 60Fe can constrain the origin of a runaway star assuming that both phenomena are caused by SN while the optically detectable nebula has already disappeared. We will implement this three-fold search strategy to identify new M7-like NSs combining these evidences. As first step, we will constrain the clusters with near-zero residual live times (which are origins of runaway stars) to most promising candidates, check those regions for gamma ray emission and use deep optical and X-ray archive observations to search for faint and fast moving blue objects (in optical) in the identified regions.

| RXJ | period [s] | kT [eV] | Mag (band) | motion [mas/yr] | distance [pc] |
|---|---|---|---|---|---|
| 1856.5-3754 | 7 | 60 | 25.7 (V) | 332 | $161^{+18}_{-14}$ |
| 0720.4-3125 | 8.39 | 86-95 | 26.6 (B) | 107.8 | $360^{+170}_{-90}$ |
| 1605.3+3249 | ? | 96 | 27.2 (B) | 145 | ? |
| 0806.4-4123 | 11.37 | 96 | >24 (B) | ? | ? |
| 1308.8+2127 | 10.31 | 55 | 28.6 | ? | ? |
| 2143.0+0654 | 9.44 | 101 | >23 (R) | ? | ? |
| 0420.0-5022 | 3.45 | 45 | 26.6 (B) | ? | ? |

Tab. 1: The magnificent seven Neutron stars. Data obtained from Haberl (2007) and references therein and Kaplan et al. (2007).

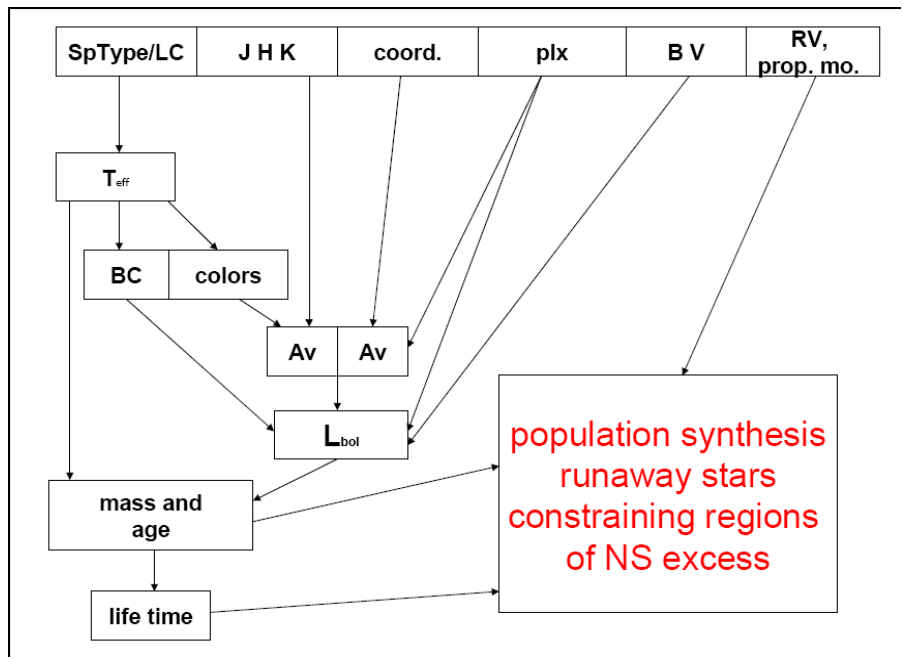

Fig. 1: Procedure of deriving mass, age and residual life time from our star sample. 3000 stars within 3 kpc have enough parameters to submit them to this procedure.

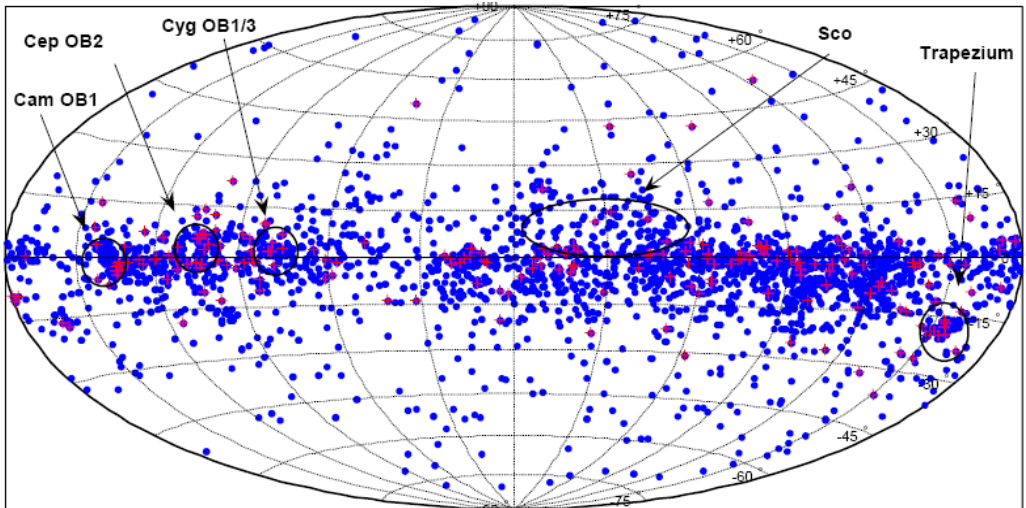

Fig. 2: All 3000 stars of our sample in galactic coordinates (blue dots). Massive stars with residual life <2Myr time are marked red. The circles show regions in the sky, which host clusters of stars with low residual life times.

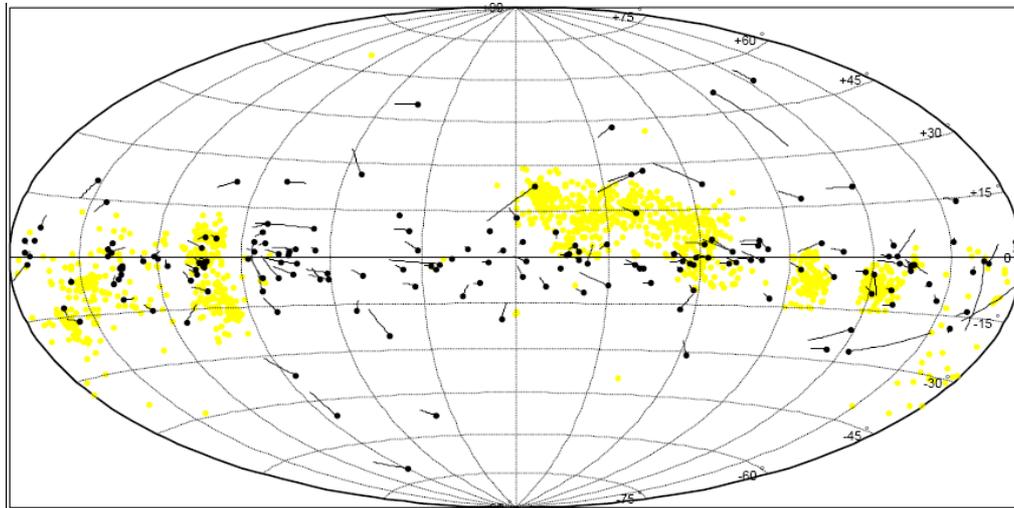

Fig. 3: Tracks of 173 runaway stars of the last 2Myrs. Background stars obtained from de Zeeuw et al. (1999) show the Gould Belt. Note that many runaway stars seem to come from OB associations marked in Fig. 2.

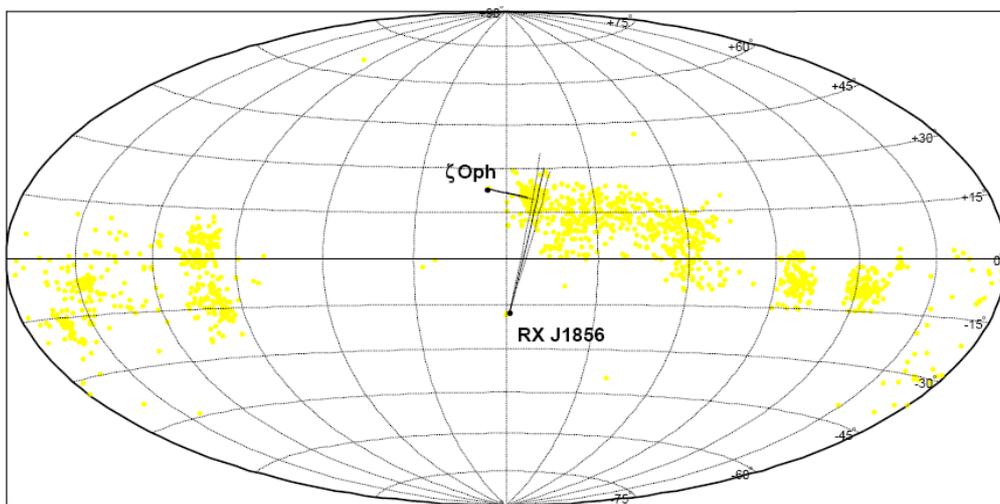

Fig. 4: As one example we show that the runaway star ζ Oph may have the same origin as RX J1856.5-3754, assuming a radial velocity of -60 km/s (approaching the sun) for RX J1856.5-3754 (distance like in Tab. 1). Tracks show motion of the last 1Myrs with their 1σ errors considering errors in proper motion and parallax.

## Acknowledgement

The work is supported by the DFG through SFB/TR7 "Gravitationswellenastronomie", project C7.